\documentstyle[11pt,paspconf,epsf]{article}

\begin{document}
\title{Point source models for the gravitational lens B1608+656:
Indeterminacy in the prediction of the Hubble constant}

\author{Gabriela Surpi and Roger Blandford}
\affil{California Institute of Technology 130-33, Pasadena CA 91125, USA}

\begin{abstract}
We apply elliptical isothermal mass
models to reproduce the point source properties, i.e.
image positions, flux density ratios and
time delay ratios,
of the quadruple lens B1608+656.
A wide set of suitable solutions is found, showing that
models that only fit the properties of point sources
are under-constrained and can lead to a large
uncertainty in the prediction of H$_\circ$. We present
two examples of models predicting H$_\circ\!=\!100{\rm
km\,s^{\!-\!1}Mpc^{\!-\!1}}$ ($\chi^2\!=\!4$)
and H$_\circ\!=\!69{\rm km\,s^{\!-\!1}Mpc^{\!-\!1}}$ ($\chi^2\!=\!24$). 
\end{abstract}


\vspace*{-0.4cm}
\section{Introduction}

Relative positions, flux ratios and time delays of the 4
images in B1608+656 have been presented here by 
Fassnacht (1999) and references therein ({\it cf} Table~\ref{obs}).
Koopmans \& Fassnacht (1999)
have concluded H$_\circ\!=\!59^{+7}_{-6}{\rm km\,s^{\!-\!1}Mpc^{\!-\!1}}$
within the context of a family of parametrized, isothermal models. Here, we
investigate whether a larger set of models allows a wider
range of Hubble constants.
%\vspace*{-0.1cm}
\section{Elliptical isothermal models}

Following Blandford \& Kundi\'c (1997),
we adopt a scaled lensing potential $\psi$ composed of two elliptical
contributions to describe the lensing galaxies G1 and G2 plus  
external shear $\gamma$:
\vspace*{-0.2cm}
\begin{equation}
\psi_= \sum_{i=1}^2 \,\, b_i\, \{s_i^2+r_i^2\, [1-e_i\, \cos (2(\varphi_i-\phi_i))]\}^{1\over 2}+
r_1^2\, \gamma\, \cos (2(\varphi_1-\varphi_\gamma))
\end{equation}
\vspace*{-0.2cm}
\normalsize

Here $(r_i,\varphi_i )$ are polar coordinates with origin at the
center of each galaxy.
$s$ measures the core radius, $e$ and $\phi$
the ellipticity and position angle of the major axis.
At large radius the mass distribution is isothermal, 
going as $\Sigma \propto r^{-1}$. 
The lenses will be fixed at $\vec{x}_{G1}\!=\!(0.446,-1.063)''$ and
$\vec{x}_{G2}\!=\!(-0.276,-0.937)''$, the centroids in H band, which
are less affected by reddening (Blandford, Surpi \& Kundi\'c 1999).

We minimize a $\chi^2$ function.
The best fit achieved, hereafter Model A, has $\chi^2\!=\!4.0$ and yields 
$H_{\circ}\!=\!100{\rm km\,s^{\!-\!1}Mpc^{\!-\!1}}$.
Models with lower values of $H_\circ$ can also be
built fixing $H_\circ$ and fitting the 3 time delays instead of the
time delay ratios. As an example we present the results of Model B
having $H_\circ\!=\!69{\rm km\,s^{\!-\!1}Mpc^{\!-\!1}}$
and $\chi^2\!=\!24.7$.
The parameters and predictions of Model A and B
are displayed in Tables~\ref{par} and~\ref{obs} respectively.
They represent reasonable mass distributions given, especially,
our ignorance of the dark matter distribution (Figure 1).

\section{Discussion}

A variety of parametrized models 
can reproduce the point source properties 
of B1608+656. This precludes an accurate determination of H$_\circ$.
To break the degeneracy, extra constraints, associated with
the extended emission of the source, have to be incorporated.
It is also helpful to specify the distribution of dark matter
on larger scale than the image distribution. A similar conclusion
has been drawn by Williams \& Saha (1999) using pixellated models.
\begin{figure}
\vspace*{-0.5cm}
\begin{center}
\leavevmode
\epsfysize=1.55in
\epsfbox{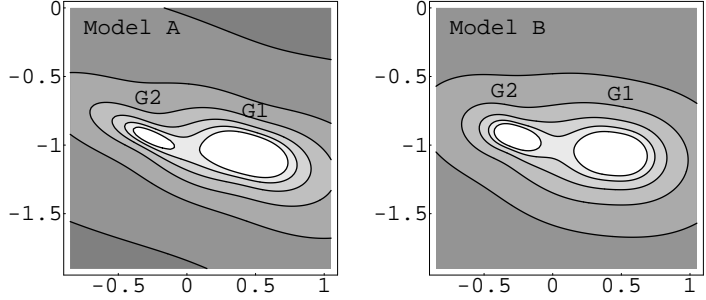}
\vspace*{-0.2cm}
\caption{Surface mass density in Models A and B}
\vspace*{-0.5cm}
\end{center}
\end{figure}

\begin{table}
\caption{Model parameters.} \label{par}
\vspace*{-0.3cm}
\begin{center}
\footnotesize
\begin{tabular}{c|cc|cc}
\tableline
&\multicolumn{2}{c|}{Model~A~(H$_\circ\!=\!100 \scriptstyle\rm km\,s^{\!-\!1}Mpc^{\!-\!1}$)}
&\multicolumn{2}{c}{Model~B~(H$_\circ\!=\!69 \scriptstyle\rm km\,s^{\!-\!1}Mpc^{\!-\!1}$)}\\
\cline{2-5}
Parameters& G1 & G2 & G1 & G2 \\
\tableline
s('') & 0.10 & 0.10       & 0.00 & 0.05 \\
b & 0.9072 & 0.2453     & 0.7797 & 0.3429 \\
e & 0.3269 & 0.6405        & 0.1570  & 0.3149 \\
$\phi(^\circ)$ & 163.45 & 154.93 & 172.62 & 160.86 \\
\tableline
$\gamma,\varphi_{\gamma}(^\circ)$ & 0.0876 & -10.92& 0.0473 & 12.47 \\
\tableline
\end{tabular}
\end{center}
\vspace*{-0.4cm}
\caption{Comparison between observations and model predictions.} \label{obs}
\vspace*{-0.3cm}
\begin{center}
\footnotesize
\begin{tabular}{cccc}
\tableline
Properties & \multicolumn{1}{c}{Observation}
& \multicolumn{1}{c}{Model A}
& \multicolumn{1}{c}{Model B} \\
\tableline
$\vec x_A('')$ & (~0.0000,~0.0000) $\pm$ (0.0023,0.0023) & (~0.0000,~0.0000) & (~0.0000,~0.0000)\\
$\vec x_B('')$ & (-0.7380,-1.9612) $\pm$ (0.0043,0.0046) & (-0.7382,-1.9613) & (-0.7365,-1.9518)\\
$\vec x_C('')$ & (-0.7446,-0.4537) $\pm$ (0.0045,0.0049) & (-0.7443,-0.4544) & (-0.7422,-0.4575)\\
$\vec x_D('')$ & (~1.1284,-1.2565) $\pm$ (0.0107,0.0124) & (~1.1271,-1.2582) & (~1.1269,-1.2207)\\
\tableline
$F_A / F_B$ & 2.042 $\pm$ 0.124 & 1.917 & 1.901 \\
$F_C / F_B$ & 1.037 $\pm$ 0.083 & 1.092 & 1.131 \\
$F_D / F_B$ & 0.350 $\pm$ 0.055 & 0.428 & 0.504 \\
\tableline
%
%$T_{AB}/T_{CB}$ & 0.79 $\pm$ 5.0 & 28.4 & 26.4 \\ 
%$T_{AB}/T_{DB}$ & 33.0 $\pm$ 5.0 & 32.0 & 30.8 \\
$T_{AB} (d)$ & 26.0 $\pm$ 5.0 & 28.4 & 27.6 \\ 
$T_{CB} (d)$ & 33.0 $\pm$ 5.0 & 32.0 & 31.7 \\
$T_{DB} (d)$ & 73.0 $\pm$ 5.0 & 68.4 & 71.3 \\
\tableline
$\chi^2$ & 0.0 & 4.0 & 24.7 \\
\tableline
\end{tabular}
\end{center}
\end{table}

\end{document}